\documentclass[11pt]{article}
\usepackage{moriond,epsfig,amsmath,amssymb}

\def\Journal#1#2#3#4{{#1} {\bf #2}, #3 (#4)}

\def\NIMA{{\em Nucl. Instrum. Methods} A}

\def\PLB{{\em Phys. Lett.}  B}

\def\PRD{{\em Phys. Rev.} D}

\def\be{\begin{equation}}
\def\ee{\end{equation}}
\def\bea{\begin{eqnarray}}
\def\eea{\end{eqnarray}}

\begin{document}
\vspace*{4cm}
\title{ELECTROWEAK PHYSICS AT HERA}

\author{ J. \v Sutiak on behalf of the H1 and ZEUS collaborations }

\address{Max-Planck-Insitut f\"ur Physik (Werner-Heisenberg-Institut),\\
F\"ohringer Ring 6, 80805 M\"unchen, Germany}

\maketitle\abstracts{Neutral-current (NC) and charged-current (CC) deep inelastic scattering (DIS) interactions have been studied in $e^+p$ and $e^-p$ collisions with longitudinally polarised lepton beams using the H1 
and ZEUS detectors at HERA.  Measurements of the polarised inclusive cross-sections are
   presented and compared   to the Standard Model (SM) expectations. The extraction of several electroweak parameters using a combined EW-QCD fit   of unpolarised data  is also discussed.
}

\section{Introduction}

Measurement of DIS of leptons on protons has been an important tool for studying the structure of the nucleon and QCD. The $ep$  collisions at the HERA  accelerator at DESY became   an essential part of this field because of the wide kinematic range covered by the machine with the  center-of-mass energy $\sqrt{s}=318\:{\rm GeV}$. 
HERA has been running since 1992 alternatively with positron and electron beams. In the period 2001-2002 the machine was upgraded to provide higher instantaneous luminosities and polarised lepton beams. The first period of running before the upgrade is called HERA~I run and the period after the upgrade HERA~II. The total integrated luminosities used in the analyses presented in this paper are summarized in  Tab.~\ref{tab:lumi}

The large amount of luminosity collected so far, together with good understanding of QCD effects and especially the availability of polarised lepton beams allow to test  the  electroweak sector of the SM. Two groups of electroweak analyses are presented in this paper. The first  group includes the results of the combined QCD-electroweak analysis of unpolarized HERA~I CC and NC cross-sections performed by the H1 collaboration (sec. \ref{sec:EWQCD}). The second group includes the results of analyses of polarised CC and NC cross-sections performed by both ZEUS and H1 (sec. \ref{sec:polar}).

\begin{table}[!ht]
\begin{center}

\begin{tabular}{|l|c|c|c|c|}

\hline 
		 &lepton&  ZEUS 	&  H1		& 	Average polarisation	\\
\hline
HERA~I& $e^+$&  $  - $	& $101\:{\rm pb}^{-1}$   	&	0	\\
		 & $e^-$&  $ - $    & $16\: {\rm pb}^{-1}$            &       0	\\
\hline 
HERA~II& $e^+$&  $30\: {\rm pb}^{-1}$  & $48\: {\rm pb}^{-1}$   &     -40\% and +33\%	\\
		  & $e^-$&  $42\: {\rm pb}^{-1}$   & $18\: {\rm pb}^{-1}$  &      -23\% and +22\%  \\
\hline
\end{tabular}

\end{center}
\caption{Integrated luminosities and average lepton beam polarisation of the data samples    used in the presented analyses.}
\label{tab:lumi}
\end{table}
\section{DIS at HERA}
\subsection{Kinematics and cross-sections}
The inclusive DIS  of leptons on nuclons can be described by 3 kinematic variables $x$, $y$ and $Q^2$. The variable $Q^2$ is the negative square of the momentum transfer $q$: $Q^2=-q^2=-(k'-k)^2$ where $k$ and $k'$ are the 4-momenta of the incoming and scattered lepton. The {\it Bjorken} variable $x$ is defined as $x=Q^2/2P\cdot q$ where $P$ is the 4-momentum  of the incoming proton. The variable $y$ defined as $y=P\cdot q/P\cdot k$ describes the inelasticity of the interaction. Depending on the type of the boson exchanged, the inclusive interactions are called either CC or NC.   

 The CC interactions   $e^{\mp}p\rightarrow \nu_e(\overline{\nu}_e)X$ are mediated by the exchange of  $W^{\mp}$ bosons. The double differential cross section for unpolarised beams at the Born level can be written as:
\be
    \frac {d^{2}\sigma_{e\pm p}^{CC}}{dxdQ^2} = \frac{G_{F}}{4\pi x} 
    \:\frac{M^{2}_{W}} {(M^{2}_{W}+Q^{2})^{2}} \tilde\sigma^{CC}_{\pm}(x,Q^2)
\label{eq:CC}
\ee
where $G_{F}$ is the Fermi constant, $M_{W}$ is the mass of the $W$ boson and $\tilde\sigma^{CC}(x,Q^2)$ is the reduced cross-section describing the structure of the proton:

\be
 \tilde\sigma^{CC}_{\pm}(x,Q^2) = Y_{+}W^{\pm}_{2}(x,Q^{2})-y^{2}W^{\pm}_{L}(x,Q^{2}) \mp Y_{-}xW^{\pm}_{3}(x,Q^2). 
\label{eq:redCC}
\ee

\noindent
The factors $Y_{\pm}$ are defined as $Y_{\pm}=1\pm\left(1-y\right)^{2}$. 
The functions $W^{\pm}_2(x,Q^2)$, $W^{\pm}_{L}(x,Q^{2})$ and $xW^{\pm}_{3}(x,Q^2)$ are  the three generalized CC proton structure functions.

The NC interaction $e^{\pm}p\rightarrow e^{\pm} X$ is mediated by the exchange of a  photon or
$Z^0$ boson. The double-differential cross-section for unpolarised beams is given by:
\be
\frac {d^{2}\sigma_{e\pm p}^{NC}}{dxdQ^2} = \frac {2\pi\alpha^2}{xQ^4}\tilde\sigma^{NC}_{\pm}(x,Q^2),
\label{eq:NC}
\ee
where $\tilde\sigma^{NC}_{\pm}(x,Q^2)$ is a combination of the generalized NC proton structure functions:
\be
\tilde\sigma^{NC}_{\pm}(x,Q^2)=Y_{+}F^{\pm}_{2}(x,Q^{2})-y^{2}F^{\pm}_{L}(x,Q^{2}) \mp Y_{-}xF^{\pm}_{3}(x,Q^2).
\label{eq:redNC}
\ee

\subsection{Event selection and reconstruction}

In CC interactions, the outgoing neutrino escapes the detection. The clear signature of these events is a hadronic final state with apparent net transverse momentum $p_T$ measured by the calorimeters. To select the CC events at HERA, the transverse momentum was required to be $p_T>12\:{\rm GeV}$. This cut implies that the CC measurement was possible only for $Q^2 \gtrsim 200\: {\rm GeV^2}$. The actual cut used by ZEUS was $200\:{\rm GeV^2}$ and $400\:{\rm GeV^2}$ for H1. The  event shape cuts were used to reject photoproduction
%  (PHP) \footnote{PHP is a NC process with very low $Q^2$. The electron is scattered at low angle and escapes down the beam-pipe. Mismeasurement of the energy can fake the missing transverse momentum.} 
  events. The events with an electron were rejected as NC background.
 the position of the event vertex and the event timing were required to be consistent with the interaction region and the beam collision timing in order to remove
 the non-$ep$ background.
 
The NC events were selected by looking for a scattered electron (positron) in the event. Both experiments developed sophisticated methods for electron-finding. The main signature used was the shower shape in the calorimeter which differents between  the electrons and  the hadrons. The electron was required to be isolated from the hadrons and to have energy $E>10\:{\rm GeV}$ at ZEUS and $E>11\:{\rm GeV}$ at H1.   
The kinematic variables for CC events were reconstructed from the hadronic system using the Jacquet-Blondel method \cite{JB}. For the NC interactions the ZEUS measurements use the double-angle method \cite{DA} which makes use of the electron scattering angle and the angle of the hadronic system. H1 use the $e\Sigma$ method \cite{eSigma} which uses the electron energy and scattering angle.

\section{Combined Electroweak-QCD Analysis of HERA~I data}\label{sec:EWQCD}
 
 The parameters of PDFs  have always been  determined with some assumptions on EW parameters. Trying to determine the EW parameters from the same data making assumptions on PDFs would be inconsistent.  The combined EW-QCD analysis therefore attempts to determine both EW and PDFs parameters in one fit to NC and CC data \cite{H1EW}. 
In this analysis,  H1 uses the NC and CC cross-sections already published by the collaboration \cite{H1_1}. 

\subsection{$W$ mass and $\sin^2 \theta_W$}
\begin{figure}[!ht]
%\rule{5cm}{0.2mm}\hfill\rule{5cm}{0.2mm}
%\vskip 2.5cm
%\rule{5cm}{0.2mm}\hfill\rule{5cm}{0.2mm}
\vfill
\begin{center}
\epsfig{figure=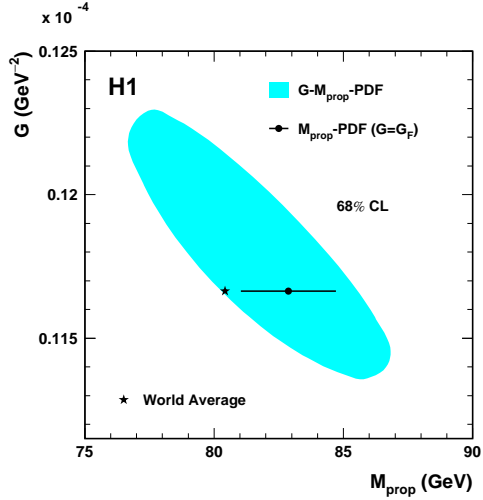,width=7.5cm}
\end{center}
\caption{The result of the fit to $G$ and $M_{prop}$. The shaded area shows the  68\% confidence level. The world average values for $M_W$ and $G_F$ are indicated with the star. The  result of the fit where  $G$ is fixed to $G_F$ is shown as a full circle with horizontal error bars.
\label{fig:MwG}}
\end{figure}

Two strategies were used to determine the mass of the W boson. 
In the first strategy, the mass of $W$ boson and the Fermi constant $G_F$ 
from Eq.~(\ref{eq:CC}) are considered to be  a propagator mass $M_{prop}$ and a normalisation constant $G$ independent of SM. These parameters were then determined from a fit to the data. The NC  data were also included in the fit to constrain the PDFs. The result  is shown in Fig.~\ref{fig:MwG}.
It is also possible to fix $G$ at $G_F$ and fit $M_{prop}$ only. The result is in this case:
\be
 M_{prop} = 82.87\pm 1.82_{\rm exp} {}^{+0.30}_{-0.16}\big|_{\rm model} \:\mathrm{ GeV}
\label{eq:Mprob}
\ee
This  is  the most accurate measurement of the CC propagator mass at HERA so far.  This result can be compared to more accurate  measurements of the $W$ propagator in a ``time-like" process at LEP and Tevatron. The world average value calculated from these measuremnets is $M_W=80.425\pm0.038\:{\rm GeV}$ \cite{PDG04}.    

 The second strategy was the analysis  in the  on-mass-shell  renormalization  scheme \cite{OMS}. Within this scheme, the constant $G_F$ in the cross-section formula~(\ref{eq:CC}) can be rewritten  in terms of the masses of $W$ and $Z$ bozons: 
\be
G_F=\frac{\pi\alpha}{\sqrt{2}M_W^2 \sin^2 \theta_W}\:\frac{1}{1-\Delta r}
=\frac{\pi\alpha}{\sqrt{2}M_W^2 \left(1-\frac{M_W^2}{M_Z^2}\right)}\frac{1}{1-\Delta r}
\label{eq:Gf}
\ee 
where $\alpha$ is the fine structure constant, $\theta_W$ is the weak mixing angle and  $\Delta r$ contains one-loop and leading higher-order EW radiative corrections. The factor $\Delta r$ is a function of the $W$, $Z$,  top quark and Higgs boson masses. Its numerical value is $\sim 0.03$. 
In this approach, also the normalization of CC cross-section depends on $M_W$ and thus provides much higher sensitivity. The combined $M_W$-PDF fit yields:
\bea
 M_W &=& 80.78\pm 0.20_{exp} {}^{+0.05}_{-0.03}\big|_{\rm model} \pm0.03_{\delta m_t} -0.08_{\delta M_H} \pm 0.03_{\delta (\Delta r)}\:{\rm GeV}\\[0.2cm]
 \sin^2\theta_W &=& 0.2151 \pm 0.0040_{exp} {}^{+0.0019}_{-0.0011}   \big|_{\rm theory}
 \label{eq:Mwsin2th}
 \eea
 
\subsection{Couplings of light quarks to $Z^0$ boson} 

The sensitivity of the NC cross-section on the quark couplings tothe  $Z^0$ boson can be seen when the proton structure functions $F_2$ and $xF_3$ are expressed in terms of the couplings and the  quark PDFs:
\bea
F^{NC}_2(x,Q^2)&=& \sum_{q} \left[ e^2_q -2e_qv_qv_e\chi_Z+(v^2_q+a^2_q)(v^2_e+a^2_e)\chi^2_Z\right] x\left[q(x,Q^2)+\bar{q}(x,Q^2)\right],
\label{eq:F2}
\\
xF^{NC}_3(x,Q^2)&=&\sum_{q} \left[-2e_qv_qv_e\chi_Z+4v_qa_qv_ea_e\chi^2_Z \right]x \left[q(x,Q^2)-\bar{q(x,Q^2)}\right],
\label{eq:xF3}
\eea
where the index $q$ runs over all the light quark flavours, $q(x,Q^2)$ is the PDF of  $q$-th flavour, $v_q$ and $a_q$ are the  vector and axial couplings of the quarks:
\bea
v_q&=&I^3_{q,L} - 2e_q\sin^2\theta_W, \\
a_q&=&I^3_{q,L},
\label{eq:qcoupl}
\eea
where $I^3_{q,L}$ is the third component of the weak isospin.  In the SM the values for $u$ quark are $a_u=1/2$, $v_u=0.196$ and for $d$ quark $a_d=-1/2$ and $v_d=-0.346$. 
The factor $\chi_Z$ is proportional to the  ratio of $Z^0$ and photon propagators:
\be
\chi_Z=\frac{G_FM^2_Z}{2\sqrt{2}\pi\alpha}\:\frac{Q^2}{Q^2+M^2_Z}
\label{eq:chiZ}
\ee  
We can see from (\ref{eq:F2}) and (\ref{eq:xF3}) that the dependence of $\chi_Z$ provides the sensitivity on both the vector and axial couplings. Moreover, the interference term allows also the determination the sign of the couplings, in contrast to LEP measurements where the couplings are measured in $Z^0$ decay \cite{LEPcoupl}.   

\begin{figure}[!ht]
\vfill
\begin{center}
\psfig{figure=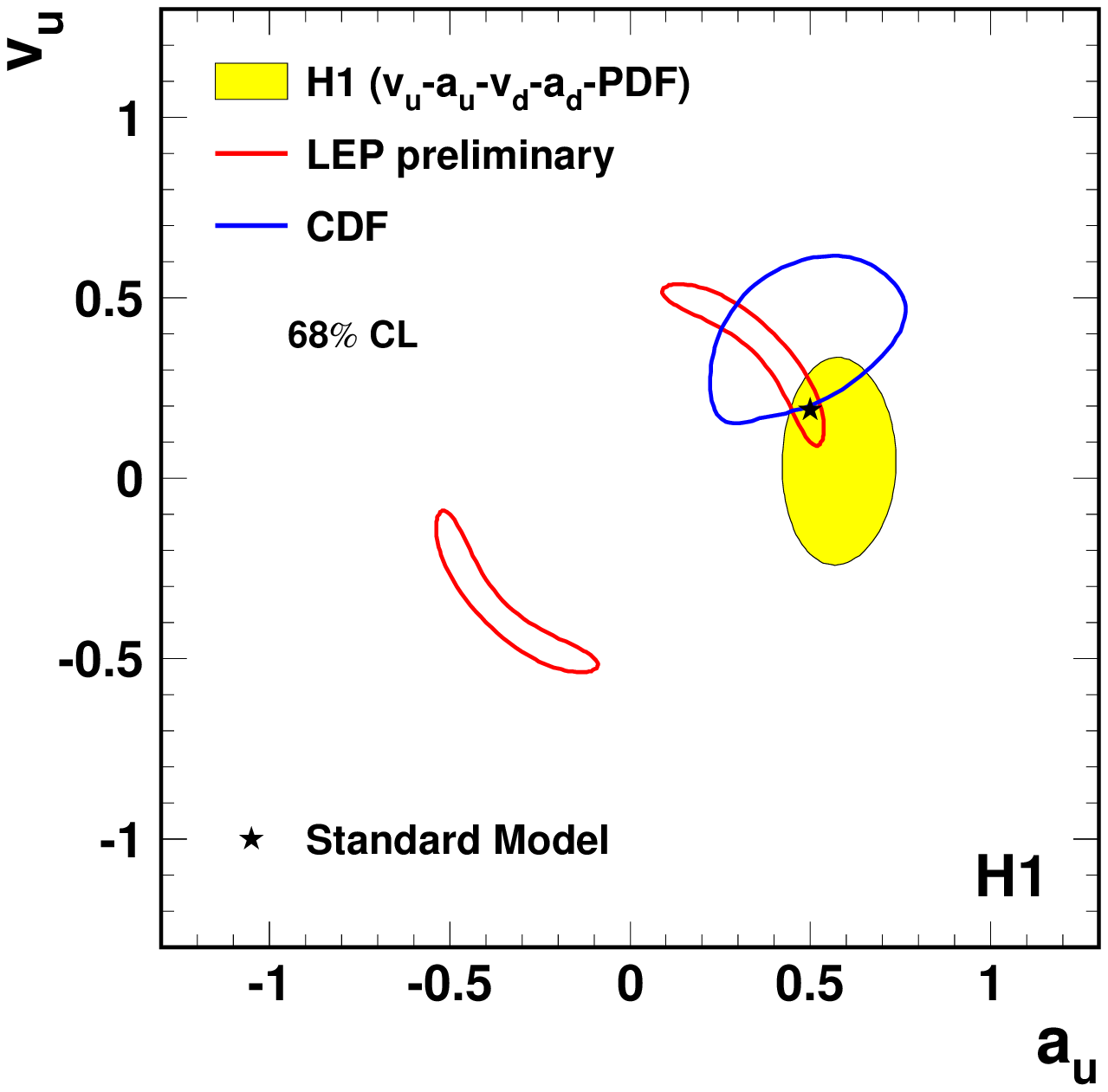,width=7.5cm}
%\hspace{0.5cm}
\psfig{figure=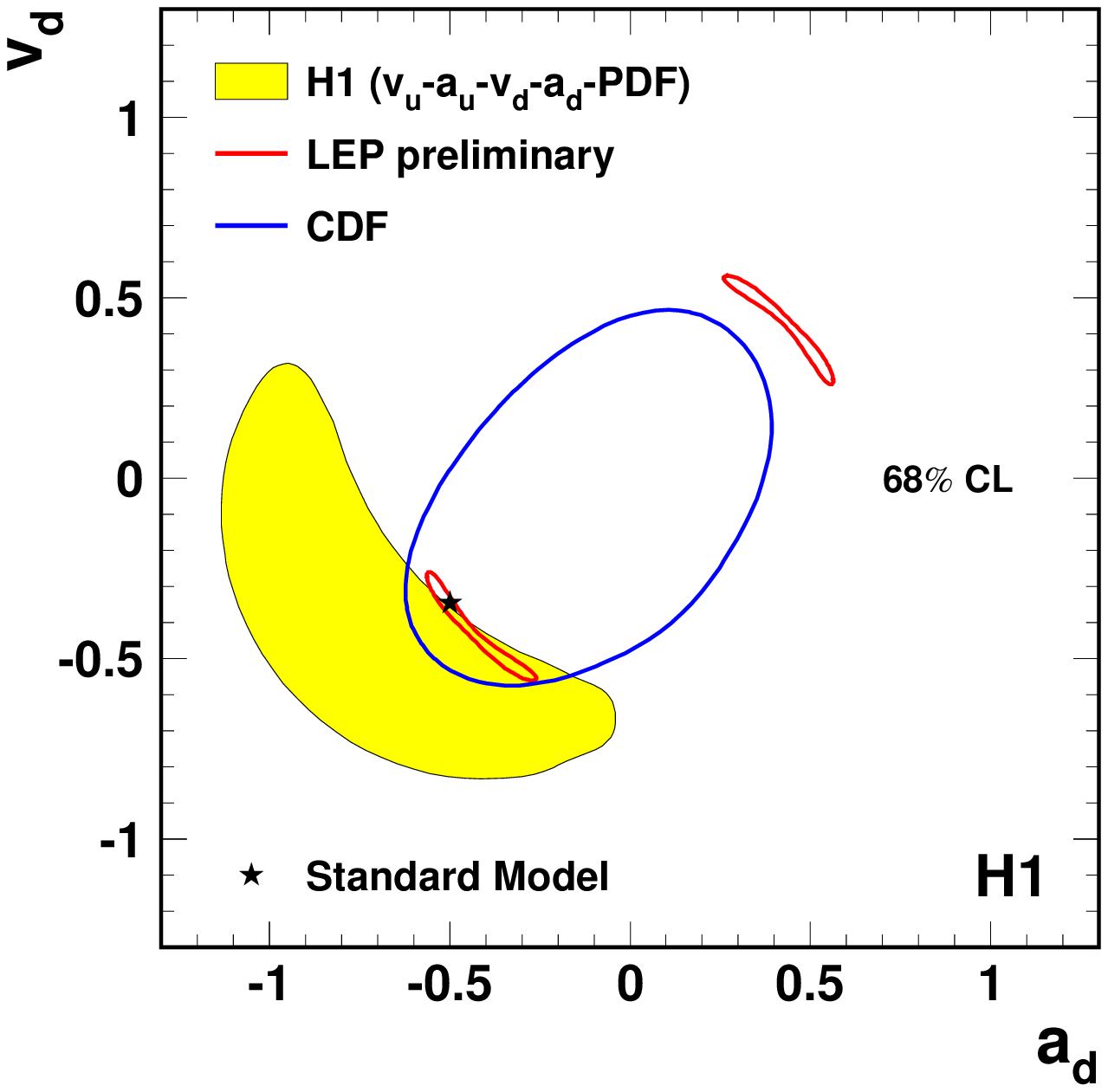,width=7.5cm}
\end{center}
\caption{The results of the fit of the weak NC couplings of $u$ (left) and $d$ (right) quarks to the $Z^0$ boson at 68\% CL (the shaded area). CDF and LEP results are shown for comparison. The SM values are plot as  stars. 
\label{fig:coupl}}
\end{figure}

The result of the fit to the data with $a_u$, $v_u$, $a_d$, $v_d$ and PDFs as free parameters is shown in Fig. \ref{fig:coupl}. The best estimates are
\begin{equation}
\begin{array}{rclcrcl}
a_u&=&0.56\pm0.10  &\quad&v_u&=&0.05\pm0.19\\
a_d&=&-0.77\pm0.37 &\quad&v_d&=&-0.50\pm0.37.
\end{array}
\label{eq:couplres}
\end{equation}
All the fitted couplings are consistent with their SM values given by Eq.~(\ref{eq:qcoupl}).

\section{Polarised cross-sections}\label{sec:polar}

In the SM only left-handed fermions and right-handed antifermions contribute to the weak interactions. The CC cross-section therefore depends linearly on lepton beam polarisation:
\be
\sigma_{e\pm p}^{CC}=(1\pm P)\sigma_{e\pm p}^{CC}(P=0),
\label{eq:polCC}
\ee
where  $\sigma_{e\pm p}^{CC}(P=0)$ is the unpolarised cross-section Eq.~(\ref{eq:CC}) and the polarisation level $P$ is defined as:
\be
P=\frac{N_R-N_L}{N_R+N_L},
\label{eq:poldef}
\ee
where $N_R$ and $N_L$ are numbers of right-  and left-handed leptons in the beam. 
Similarly we can express the polarisation dependence of NC cross-section as:
\be
\frac{d^2\sigma^{NC}_{e^{\pm}p}}{dxdQ^2}=\frac{2\pi\alpha^2}{xQ^2}\left[ H^{\pm}_0 + PH^{\pm}_p\right],
\label{eq:polNC}
\ee
where the structure function term is split into unpolarised ($H^{\pm}_0$) and polarised ($H^{\pm}_p$) part. The dependence on polarisation comes from the contribution of $Z^0$ exchange which increases with $Q^2$. On the contrary, in case of CC interaction, the polarisation dependance does not change with $Q^2$.

\begin{figure}[!ht]
\vfill
\begin{center}
\psfig{figure=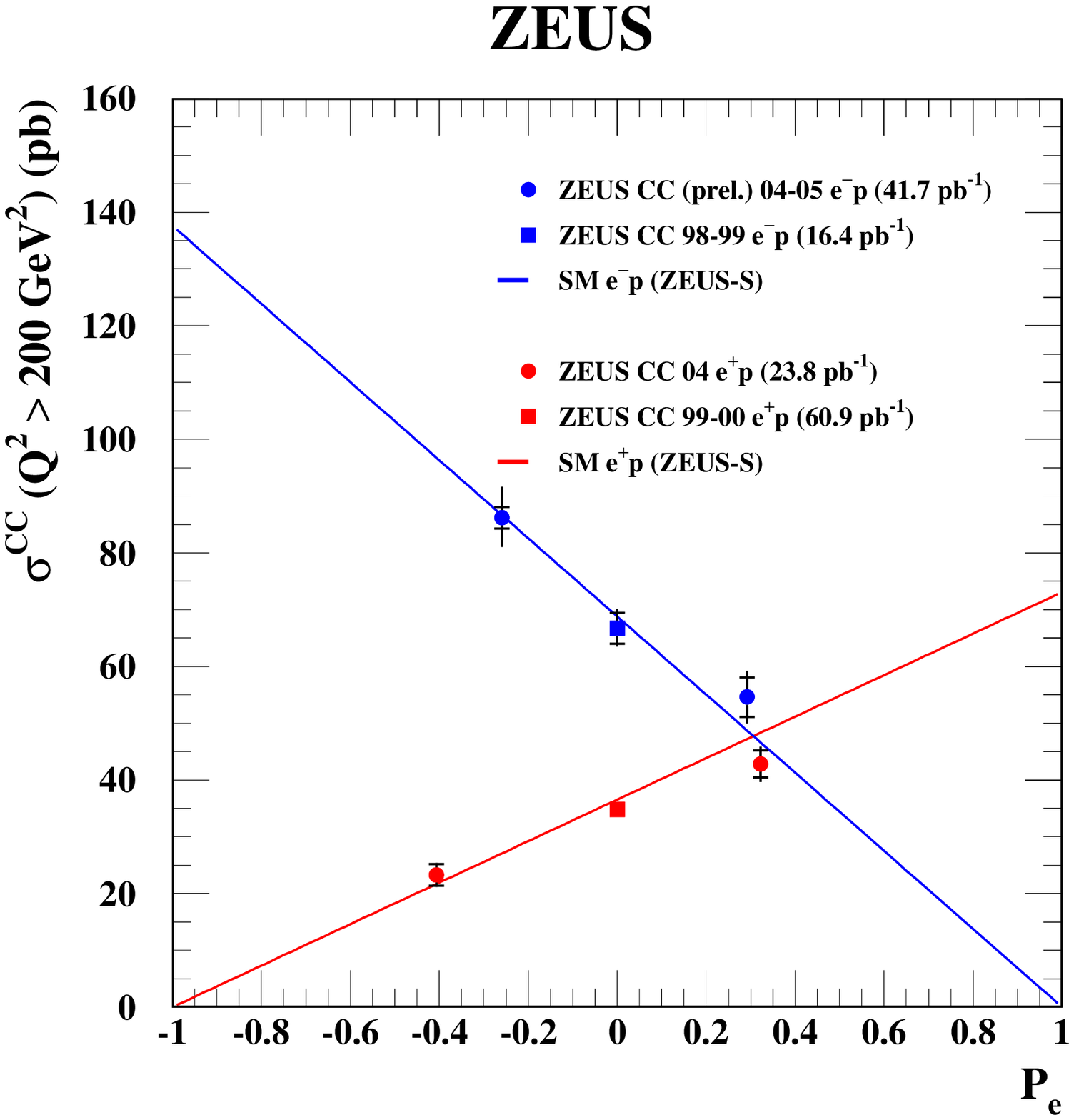,width=7.5cm}
%\hspace{0.5cm}
\psfig{figure=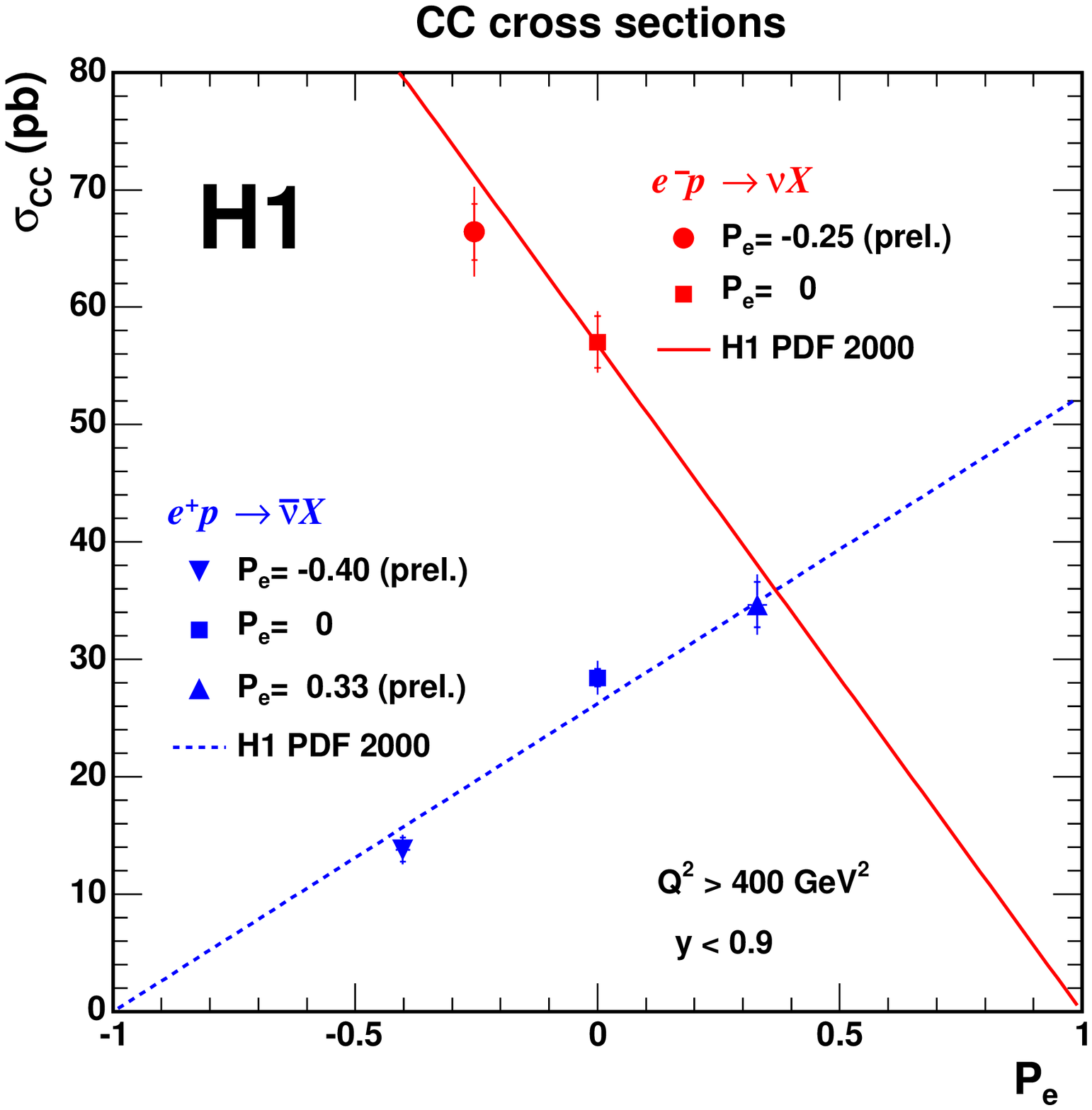,width=7.cm,height=7.cm}
\end{center}
\caption{The total CC cross-section as a function of lepton beam polarisation measured by the ZEUS (left) and H1 (right) experiments. The points are the measured values and the lines represent SM expectations. 
\label{fig:CCvsPol}}
\end{figure}

The total CC cross-section has been measured as a function of lepton beam polarisation by both ZEUS and H1 collaboration. ZEUS measured the cross-section in the kinematic range $Q^2>200\:{\rm GeV^2}$ for the both lepton charges, each at 2 values of average polarisation. H1 measured the CC cross-section in the kinamatic range $Q^2>400\:{\rm GeV^2}$ and $y<0.9$. The measurement was performed at two values of polarisation for positrons and at one value of polarisation for electrons. The results are presented in Fig. \ref{fig:CCvsPol}. The measurements are consistent with unpolarised HERA~I cross-sections \cite{ZEUSunpol,H1unpol}  and they  are in perfect agreement with the SM at both experiments. Both collaborations have published the first measurements with a positron beam \cite{H1pol,ZEUSpol}.  
The extrapolation of ZEUS data to $P=-1$ for positron cross-sections yields:
\be
\sigma^{CC}_{+}(P=-1)=7.4\pm4.0_{\rm stat}\pm1.2_{\rm sys}\:{\rm pb}.
\label{eq:extZEUS}
\ee
The extrapolation of H1 data gives:
\be
\sigma^{CC}_{+}(P=-1)=-3.9\pm2.3_{\rm stat}\pm0.7_{\rm sys}\pm0.8_{\rm pol}\:{\rm pb}.
\label{eq:extH1}
\ee
Both extrapolations are consistent with zero as expected by the SM. The deviation is less than $2$  standard deviations for ZEUS and  $1.6$ for H1.

\begin{figure}[!ht]
 \vfill
 \begin{center}
 \psfig{figure=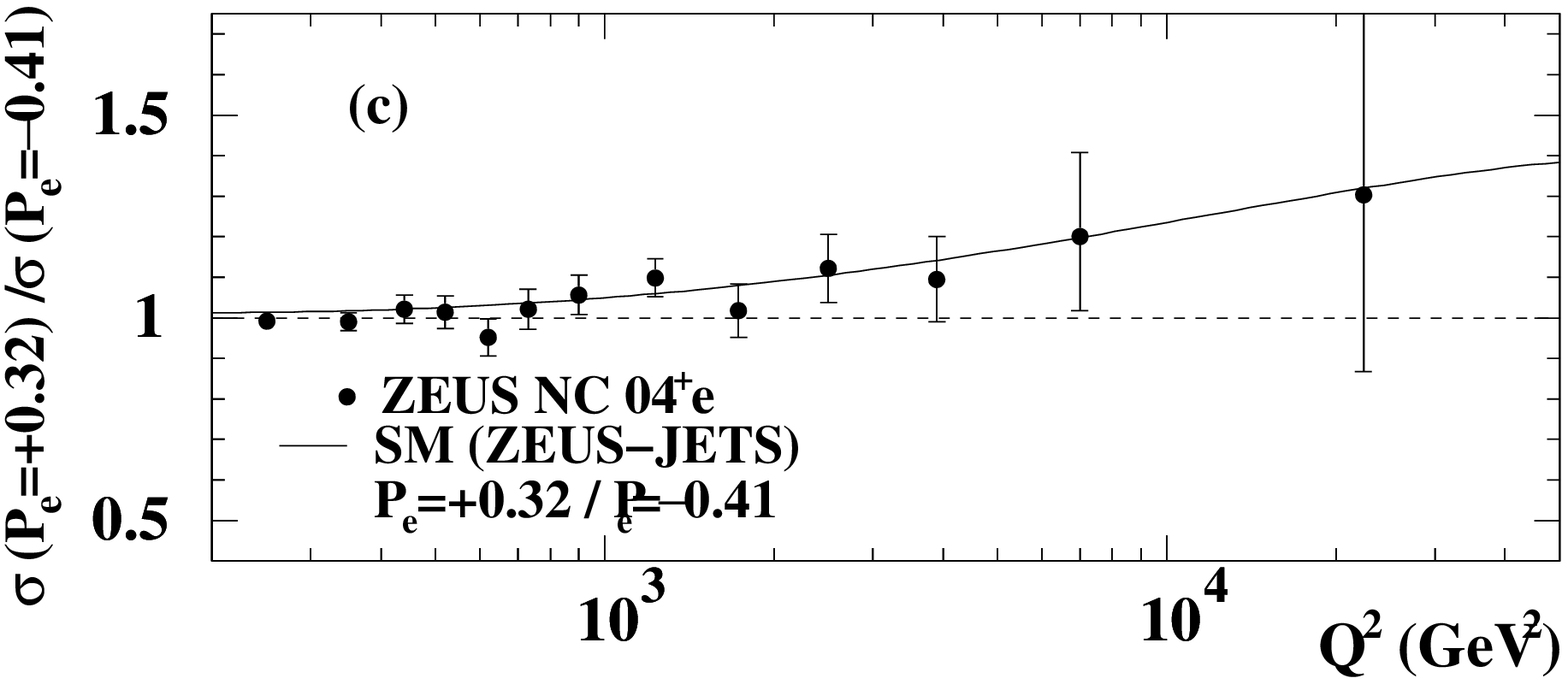,width=10cm} 
 \vspace{0.5cm}
 \psfig{figure=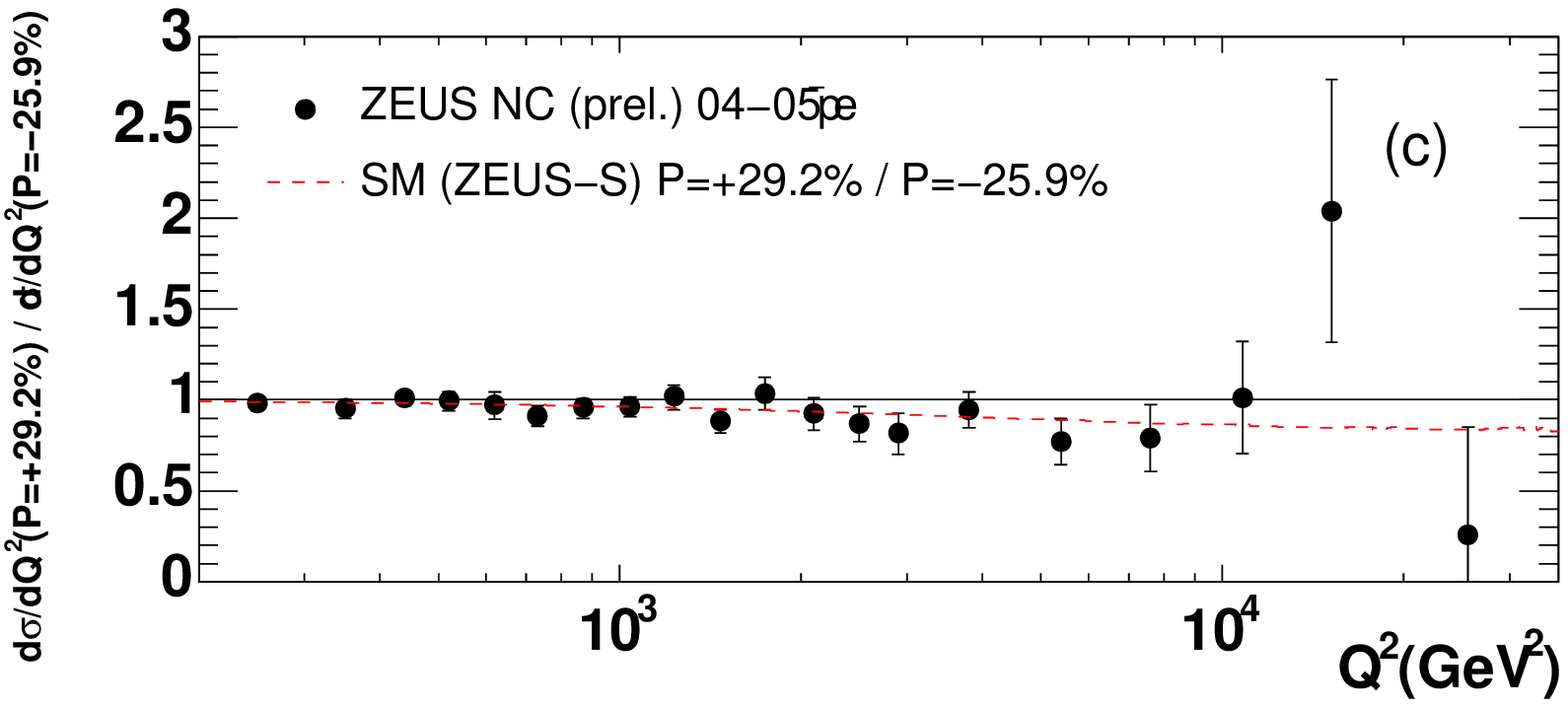,width=10cm}
\end{center}
\caption{The ratio of NC cross-section at positive polarisation over the cross-section at negative polarisation as a function of $Q^2$. The top plot shows the positron data and the bottom plot the electron data. The solid line in the top plot and the dashed line in the bottom plot show the SM expectation. 
\label{fig:NCvsPol}}
\end{figure}

The effect of polarisation on NC cross-section at  $Q^2>200\:{\rm GeV^2}$ was studied by ZEUS. The result is presented in Fig. \ref{fig:NCvsPol}. The ratio of the cross-section for positive polarisation over the cross-section for negative polarisation is plotted as a function of $Q^2$ for both positrons and electrons. The data are compared to SM model expectation and to a case with no effect of polarisation. The data seem to be more consistent with the  SM case although the statistics are too small to make any strong conclusions.

\end{document}